\documentclass[12 pt]{article}
\usepackage[english]{babel}
\usepackage{amssymb}
\usepackage{inputenc}
\newcommand{\bc}{\begin{center}}
\newcommand{\ec}{\end{center}}
\newcommand{\be}{\begin{equation}}
\newcommand{\ee}{\end{equation}}
\newcommand{\bea}{\begin{eqnarray}}
\newcommand{\eea}{\end{eqnarray}}
\newcommand{\ba}{\begin{array}}
\newcommand{\ea}{\end{array}}
\newcommand{\edc}{\end{document}}

\def\O{\Omega}

\def\s{\sigma}

\def\l{\lambda}

\begin{document}

\begin{center}
{\bf FERTILE THREE STATE HARD-CORE MODELS ON A CAYLEY TREE}\\
\vspace{0.3cm} {\bf U.A. Rozikov$^{1,2},$ Sh.A. Shoyusupov$^1$}

$^1${\it Institute of Mathematics, Tashkent, Uzbekistan.}
\\ e-mail: rozikovu@yandex.ru,
shokir79@mail.ru\\

$^2${\it School of Mathematical Sciences, GC University, Lahore,
Pakistan.}
\end{center}

{\bf Abstract.} We consider  nearest-neighbor fertile hard-core
models, with three states , on a homogeneous Cayley tree. It is
known that there are four type of such models. We investigate all of
them and describe translation-invariant and periodic hard-core Gibbs
measures. Also we construct a continuum set of non-periodic Gibbs
measures.

\section{Introduction}

\qquad A Cayley tree $T^k=(V,L)$ of order $k\geq 1$ is defined as an
infinite homogeneous tree, i.e., a graph without cycles, with
exactly $k+1$ edges incident to each site. Here $V$ is the set of
sites and $L$ is the set of edges. Fix a site $x^0$ (the origin) and
set: $V_n=\{x\in V:$ dist $(x^0,x) \leq n\}$, $W_n=\{x\in V:$ dist
$(x^0,x) = n\}$, where the distance between  $x$, $y\in V$ is the
number of edges in the shortest path $x\to y$.

We consider  nearest-neighbor hard-core models, with three states ,
on a homogeneous Cayley tree. In these models one assigns, to each
site $x$, values $\sigma (x)\in \{0,1,2\}$. Values $\sigma (x)=1,2$
mean that site $x$ is `occupied' and $\sigma (x)=0$ that $x$ is
`vacant'.

A configuration $\sigma$ on the tree is a collection
$\{\sigma(x),$ $x\in V\}$ considered also as a function $V\to
\{0,1,2\}$. In a similar fashion one defines a configuration in
$V_n$ and $W_n$.

In this paper we consider the fertile graphs (see [2], p.248) with
three vertices $0,1,2$ (on the set of values $\s(x)$), with edges
and loops as follows:

the {\it "wrench"}: $\{0,1\}, \{0,2\};$ loops at 0 and 1;

the {\it "wand"}: $\{0,1\}, \{0,2\};$ loops at 1 and 2;

the {\it "hinge"}: $\{0,1\}, \{0,2\};$ loops at 0, 1 and 2;

the {\it "pipe"}: $\{0,1\}, \{1,2\};$ loop at 0.

Denote $O=\{wrench, wand, hinge, pipe\}.$

Another graph which is non fertile is called sterile (see [2],
p.247).

For $G\in O$ we call $\sigma$ a $G-$admissible  configuration (on
the tree, in $V_n$ or $W_n$) if $\{\sigma (x),\sigma (y)\}$ is an
edge of $G$ $\forall$ nearest-neighbor pair $x,y$ (from $V$, $V_n$
or $W_n$, respectively). Denote the set of $G-$admissible
configurations by $\Omega^G$ ($\Omega_{V_n}^G$ and
$\Omega_{W_n}^G$).

A set of activities (see [2]) for a graph $G$ is a function $\l:G
\to R_+$ from the vertices of $G$ to the positive reals. The value
$\l_i$ of $\l$ at a vertex $i\in\{0,1,2\}$ is called its "activity".

For a given $G$ and $\l$ we define Hamiltonian of the $(G-)$ hard
core model as
$$ H^\l_G(\s)=\left\{\begin{array}{ll}
\sum_{x\in V}\ln \l_{\s(x)},\ \ \mbox{if}\ \ \s\in \O^G,\\
+\infty \ \ \mbox{otherwise}.
\end{array}\right.\eqno(1.1)$$

 The hard-core model is interesting from the point of view of
statistical mechanics, as well of combinatorics and the theory of
neuron networks [3], [5].

Let ${\mathbf B}$ be the sigma-algebra generated by the cylinder
subsets of $\Omega^G$. Furthermore, $\forall$ $n$, ${\mathbf
B}_{V_n}$ stands for the sub-algebra of ${\mathbf B}$ generated by
events $\{\sigma\in\Omega^G$: $\sigma\big|_{V_n}=\sigma_n\}$ where
$\sigma_n$: $x\in V_n\mapsto\sigma_n (x)$ is an admissible
configuration in $V_n$ and $\sigma\big|_{V_n}$ the restriction of
$\sigma$ on $V_n$.

{\bf Definition 1.} A (three state) $G-$hard core Gibbs measure is a
probability measure $\mu$ on $(\Omega^G,{\mathbf B})$ such that,
$\forall$ $n$ and $\sigma_n\in\Omega_{V_n}^G$:

$$
\mu\left\{\sigma\in\Omega^G:\;\sigma\big|_{V_n}=\sigma_n\right\}
=\int_{\Omega^G}\mu
(d\omega)P_n\big(\sigma_n\,\big|\,\omega_{W_{n+1}} \big), \eqno(1.2)
$$
where
$$
P_n\big(\sigma_n\,\big|\,\omega_{W_{n+1}}
\big)=\frac{\exp(-H^\l_G(\s_n))}{Z_n (\lambda ;\omega |_{W_{n+1}})}
{\mathbf 1}\left(\sigma_n
\vee\omega\big|_{W_{n+1}}\in\Omega^G_{V_{n+1}}\right).
$$

Symbol $\vee$ means concatenation of configurations and
$Z_n\left(\lambda;\omega\big|_{W_{n+1}}\right)$ is the partition
function with the boundary condition $\omega\big|_{W_n}$:
$$
Z_n\left(\lambda;\omega\big|_{W_{n+1}}\right)
=\sum_{{\widetilde\sigma}_n\in\Omega^G_{V_n}}\exp(-H^\l_G(\s_n)){\mathbf
1} \left({\widetilde\sigma}_n\vee\omega\big|_{W_{n+1}}
\in\Omega^G_{V_{n+1}}\right).\eqno(1.3)
$$

In [2] it was proven that (i) for every sterile graph $G$ and any
positive activity set on $G$ there is a unique invariant Gibbs
measure on $\O^G$; (ii) for any fertile graph $G$ there is a set of
activities $\l$ on $G$ for which $\O^G$ has at least two simple,
invariant Gibbs measures.

In this paper we shall consider the case $\l_0=1,$ $\l_1=\l_2=\l>0$
and describe corresponding translation-invariant, periodic and some
non-periodic Gibbs measures. In [7] these problems were solved for
$G=$wrench case. So we shall consider cases hinge, pipe and wand.
Our some results improve the analogical results of [7].

The paper is organized as follows. In section 2 we reduce our
problems to solve a system of functional equations which depends on
adjacency matrix of $G\in O.$ Section 3 is devoted to
translation-invariant Gibbs measures. Sections 4 and 5 are devoted
to periodic and non-periodic Gibbs measures respectively. All
sections contain some remarks which compare our results with known
results.

\section{System of functional equations}

\qquad Write $x<y$ if the path from $x^0$ to $y$ goes through $y$.
Call vertex $y$ a direct successor of $x$ if $y>x$ and $x,y$ are
nearest neighbors. Denote by $S(x)$ the set of direct successors
of $x$. Note that any vertex $x\ne x^0$ has $k$ direct successors
and $x^0$ has $k+1$.

 For
$\sigma_n\in\Omega_{V_n}^G$ we define : $\#\sigma_n$ $=$
$\sum\limits_{x\in V_n}{\mathbf 1}(\sigma_n(x)\geq 1)$ (the number
of occupied sites in $\sigma_n$).

Let $z:\;x\mapsto z_x=(z_{0,x}, z_{1,x}, z_{2,x}) \in{\mathbf
R}^3_+$ be a vector-valued function on $V$. Given $n=1,2,\ldots$,
and $\l>0$ consider the probability distribution $\mu^{(n)}$ on
$\Omega_{V_n}^G$ defined by
$$
\mu^{(n)}(\sigma_n)=\frac{1}{Z_n}\lambda^{\#\sigma_n} \prod_{x\in
W_n}z_{\sigma(x),x}.\eqno(2.1)
$$

Here  $Z_n$ is the corresponding partition function:
$$
Z_n=\sum_{{\widetilde\sigma}_n\in\Omega^G_{V_n}}
\lambda^{\#{\widetilde\sigma}_n}\prod_{x\in W_n}
z_{{\widetilde\sigma}(x),x}.
$$

We say that the probability distributions $\mu^{(n)}$ are compatible
if $\forall$ $n\geq 1$ and $\sigma_{n-1}\in\Omega^G_{V_{n-1}}$:

$$
\sum_{\omega_n\in\Omega^G_{W_n}}
\mu^{(n)}(\sigma_{n-1}\vee\omega_n){\mathbf 1}(
\sigma_{n-1}\vee\omega_n\in\Omega^G_{V_n})=
\mu^{(n-1)}(\sigma_{n-1}).\eqno(2.2)
$$
In this case there exists a unique probability measure $\mu$ on
$(\Omega^G ,{\mathbf B})$ such that, $\forall$ $n$ and
$\sigma_n\in\Omega^G_{V_n}$, $\mu \left(\left\{\sigma
\Big|_{V_n}=\sigma_n\right\}\right)=\mu^{(n)}(\sigma_n)$.\vskip 0,3
truecm

{\bf Definition 2.} Measure $\mu$ defined by (2.1), (2.2) is called
a $(G-)$hard core Gibbs measure with $\lambda>0$, corresponding to
function $z:\,x\in V \setminus\{x^0\}\mapsto z_x$. The set of such
measures (for all possible choices of $z$) is denoted by ${\cal
S}_G$.

For graph $G$ denote by $L(G)$ the set of its edges and by $A\equiv
A^G=\big(a_{ij}\big)_{i,j=0,1,2}$ the adjacency matrix of $G$ i.e.
$$ a_{ij}\equiv a^G_{ij}=\left\{\begin{array}{ll}
1,\ \ \mbox{if}\ \ \{i,j\}\in L(G),\\
0 \ \ \mbox{otherwise}.
\end{array}\right.$$

The following statement describes conditions on $z_x$ guaranteeing
compatibility of distributions $\mu^{(n)}$. \vskip 0,3 truecm

{\bf Theorem 1. }  {\it Probability distributions $\mu^{(n)}$,
$n=1,2,\ldots$, in (2.1) are compatible iff for any $x\in V$ the
following system of equations holds:
$$
z'_{1,x}=\lambda \prod_{y\in S(x)}{a_{10}+
a_{11}z'_{1,y}+a_{12}z'_{2,y}\over
a_{00}+a_{01}z'_{1,y}+a_{02}z'_{2,y}},
$$
$$
z'_{2,x}=\lambda \prod_{y\in S(x)}{a_{20}+
a_{21}z'_{1,y}+a_{22}z'_{2,y}\over
a_{00}+a_{01}z'_{1,y}+a_{02}z'_{2,y}}, \eqno(2.3)$$ where
$z'_{i,x}=\lambda z_{i,x}/z_{0,x}, \ \ i=1,2$.} \vskip 0,5 truecm

{\bf Proof.} Left hand side of (2.2) can be written as:
$$
\frac{1}{Z_n}\lambda^{\#\sigma_{n-1}} \prod_{x\in
W_{n-1}}\prod_{y\in S(x)}
\left(a_{\s_{n-1}(x)0}z_{0,y}+a_{\sigma_{n-1}(x)1}\lambda
z_{1,y}+a_{\sigma_{n-1}(x)2}\lambda z_{2,y}\right).\eqno(2.4)
$$
{\sl Sufficiency}. Suppose that (2.3) holds. It is equivalent to
the representations
$$
\prod_{y\in S(x)}\left(a_{i0}z_{0,y}+\lambda a_{i1} z_{1,y}+\lambda
a_{i2}z_{2,y}\right) =a(x)z_{i,x},\ \ i=0,1,2. \eqno(2.5)
$$
for some function $a(x)>0$, $x\in V$. Setting
$A_n=\prod\limits_{x\in W_n} a(x)$ and substituting (2.1) into LHS
of  (2.2), we get (2.4) and by (2.5) we have
$$
\frac{1}{Z_n} \lambda^{\#\sigma_{n-1}}\prod_{x\in W_{n-1}}
z_{\sigma_{n-1}(x),x}a(x)
=\frac{A_{n-1}}{Z_n}\lambda^{\#\sigma_{n-1}} \prod_{x\in
W_{n-1}}z_{\sigma_{n-1}(x),x}.
$$

We should have
$$
\sum_{\sigma_{n-1}\in\Omega^G_{V_{n-1}}}
\sum_{\omega_n\in\Omega^G_{W_n}}{\mathbf 1}
\left(\sigma_{n-1}\vee\omega_n\in\Omega^G_{V_n}\right)
\mu^{(n)}(\sigma_{n-1}\vee\omega_n)=1.
$$
hence $A_{n-1}/Z_n$ $=$ $1/Z_{n-1}$, and (2.2) holds.

{\sl Necessity}. Suppose that (2.2) holds; we want to prove (2.3).
Substituting (2.1) in (2.2) and using (2.4), we obtain that
$\forall$ $\sigma_{n-1}\in\Omega^G_{V_{n-1}}$:
$$
\frac{1}{Z_n}\lambda^{\#\sigma_{n-1}} \prod_{x\in
W_{n-1}}\prod_{y\in S(x)}
\left(a_{\s_{n-1}(x)0}z_{0,y}+a_{\sigma_{n-1}(x)1}\lambda
z_{1,y}+a_{\sigma_{n-1}(x)2}\lambda z_{2,y}\right)= \prod_{x\in
W_{n-1}}z_{\sigma_{n-1}(x),x}.
$$

From this equality follows
$$
\frac{Z_{n-1}}{Z_n} \prod_{x\in W_{n-1}}\prod_{y\in
S(x)}\left(a_{i0}z_{0,y}+\lambda a_{i1} z_{1,y}+\lambda a_{i2}
z_{2,y}\right) =\prod_{x\in W_{n-1}}z_{i,x},\ \ i=0,1,2. \eqno(2.6)
$$
Denoting $z'_{i,x}=\lambda z_{i,x}/z_{0,x}, \ \ i=1,2$ from (2.6) we
get (2.3). \hfill$\blacksquare$\vskip 0,3 truecm

{\bf Remark 1.} For $G=$wrench from Theorem 1 one gets Theorem 1 of
[7].\vskip 0,3 truecm

{\bf Remark 2.} One can similarly prove Theorem 1 for very general
setting: $\s(x)$ takes values $0,1,...,q;$ $G$ is a fixed graph with
$q\geq 1$ vertices; $\l:i\in G\to \l_i\in R_+$ is a given function.
Then (2.3) has the form
$$ z_{j,x}={\l_j\over \l_0}\prod_{y\in S(x)}
\frac{a_{j0}+\sum^q_{i=1}a_{ji}z_{i,y}}{a_{00}+\sum^q_{i=1}a_{0i}z_{i,y}},\
\ j=1,2,...q.$$ However, the analysis of solutions of the equation
is very difficult.

\section{Translation-invariant Gibbs measures}

\qquad We set in future $z_{0,x}\equiv 1$ and
$z_{i,x}=z'_{i,x}>0,\ \ i=1,2$. Then $\forall$ function $x\in
V\mapsto z_x=(z_{1,x},z_{2,x})$ satisfying
$$
z_{i,x}=\lambda \prod_{y\in S(x)}{a_{i0}+
a_{i1}z_{1,y}+a_{i2}z_{2,y}\over
a_{00}+a_{01}z_{1,y}+a_{02}z_{2,y}}, \ \ i=1,2\eqno(3.1)
$$
there exists a unique $G-$hard core Gibbs measure $\mu$ and vice
versa. It is natural to begin with translation-invariant solutions
where $z_x=z$ is constant.

\subsection{Case hinge}
 In this case assuming $z_x=z$ we obtain from (3.1) the following system of
equations:

$$ \left\{\begin{array}{ll}
z_1=\lambda\bigg({1+ z_1\over 1+z_1+z_2}\bigg)^k,\\
z_2=\lambda\bigg({1+z_2\over 1+z_1+z_2}\bigg)^k.
\end{array}\right.\eqno(3.2)$$

Subtracting from the first equation of system (3.2) the second one
we get
$$
(z_1-z_2)\left[1-\lambda
\frac{(1+z_1)^{k-1}+...+(1+z_2)^{k-1}}{(1+z_1+z_2)^k}\right]=0.$$
Consequently, we have $z_1=z_2$ and
$$
(1+z_1+z_2)^k=\lambda
\big((1+z_1)^{k-1}+...+(1+z_2)^{k-1}\big),\eqno(3.3)
$$ if $z_1\ne z_2.$ For $z_1=z_2=z$ from system (3.2) we have
$$
\lambda^{-1}z=f(z)=\bigg({1+z\over 1+2z}\bigg)^k.\eqno(3.4)
$$

The function $f(z)$ is decreasing for $z>0$ which implies that
equation (3.4) has unique solution $z^*=z^*(k, \lambda)$ for any
$\lambda>0.$

If (3.3) is satisfied then we assume $k=2$ and from (3.3) we have
$$
1+z_1+z_2={\lambda+\sqrt{\lambda^2+4\lambda}\over 2}. \eqno(3.5)
$$
Using this equality from first equation of the system (3.2) we
have for $k=2$
$$
z^{(1)}_1=\bigg({1+\sqrt{1-4a^2}\over 2a}\bigg)^2, \ \
z^{(2)}_1=\bigg({1-\sqrt{1-4a^2}\over 2a}\bigg)^2, \eqno(3.6)
$$ if
$\lambda >9/4$ where $a=2(\sqrt{\lambda}+\sqrt{\lambda+4})^{-1}.$
Using the second equation we also have $z^{(1)}_2, z^{(2)}_2\in
\{z^{(1)}_1, z^{(2)}_1\}.$ Since $z_1\ne z_2$ we conclude that
$z_1=z^{(1)}_1, z_2=z^{(2)}_1$ and  $z_1=z^{(2)}_1,
z_2=z^{(1)}_1.$ It is easy to check that these solutions satisfies
the condition (3.5).

Thus if $k=2$, $\lambda> {9\over 4}$ then the system (3.2) has three
solutions $(z^*,z^*), (z^{(1)}_1, z^{(2)}_1), (z^{(2)}_1,
z^{(1)}_1),$ where $z^*$ is the unique solution of (3.4) and
$z^{(i)}_1$, $i=1,2$ is defined in (3.6). Note that
$z^{(1)}_1={1\over z^{(2)}_2}.$ Consequently by Theorem 1 we get the
following \vskip 0,5 truecm

{\bf Theorem 2.} {\it If $k=2$ then for the hinge case

1) for $\lambda\leq {9\over 4}$ there exists unique hard-core
translation-invariant Gibbs measure $\mu_0$; \

2) for $\lambda> {9\over 4}$ there are at least three hard-core
translation-invariant Gibbs measures $\mu_i,$ $i=0,1,2$.} \vskip 0,5
truecm

{\bf Remark 3.} 1) Note that the idea of analysis of solutions (3.2)
is taken from [7].\\ 2) The value $\lambda= \lambda_{\rm cr}={9\over
4}$ is exactly the critical value for $k=2.$ Clearly  $\lambda_{\rm
cr}<4=\lambda^{\rm HC}_{\rm cr}$ for $k=2.$ Here $\lambda^{\rm
HC}_{\rm cr}={1\over k-1}({k\over k-1})^k$ is the
critical value for two state hard-core model [10]. \\

{\bf Proposition 1.} {\it If $z_x=(z_{1,x}, z_{2,x})$ is a solution
of (3.1) in the case hinge then $z^-_i\leq z_{i,x} \leq z^+_i, \ \
\mbox{for any} \ \ i=1, 2, \ \ x\in V,$ where $(z^-_1, z^+_1, z^-_2,
z^+_2)$ is a solution of}
$$
\left\{\begin{array}{llll}
z^-_1=\lambda\bigg({1+z^-_1\over 1+z^-_1+z^+_2}\bigg)^k,\\
z^+_1=\lambda\bigg({1+z^+_1\over 1+z^+_1+z^-_2}\bigg)^k,\\
z^-_2=\lambda\bigg({1+z^-_2\over 1+z^+_1+z^-_2}\bigg)^k,\\
z^+_2=\lambda\bigg({1+z^+_2\over 1+z^-_1+z^+_2}\bigg)^k.
\end{array}\right.\eqno(3.7)
$$

{\bf Proof.} Is very similar to proof of Proposition 5 [7]. \vskip
0,5 truecm

{\bf Proposition 2.} {\it If $z=(z_1^-, z^+_1, z^-_2, z^+_2)$ a
solution of (3.7) then $z^-_1=z^+_1$ iff $z^-_2=z^+_2$.} \vskip
0,5 truecm

{\bf Proof.} See [7], Proposition 6. \vskip 0,3 truecm

{\bf Corollary 1.} If the system (3.7) has unique solution then
system (3.1) also has unique solution. Moreover this solution is
$z_x=(z^*_1,z^*_2), \ \ x\in V$ where $(z^*_1,z^*_2)$ is the
unique solution of (3.2).

Now we shall find exact values of $z^-_i, z^+_i, i=1,2$ for $k=2.$

Consider the system consisting of the first and the last equations
of (3.7):
$$
\left\{\begin{array}{ll}
z^-_1=\lambda\bigg({1+z^-_1\over 1+z^-_1+z^+_2}\bigg)^k,\\
z^+_2=\lambda\bigg({1+z^+_2\over 1+z^-_1+z^+_2}\bigg)^k.\\
\end{array}\right.\eqno(3.8)
$$

If $z^-_1=z^+_2$ then this system has unique solution. In case
$z^-_1\neq z^+_2$ we get
$$
(1+z^-_1+z^+_2)^k=\lambda
\big((1+z^-_1)^{k-1}+...+(1+z^+_2)^{k-1}\big).\eqno(3.9)
$$
If $k=2$ then from (3.9) we have
$$
1+z^-_1+z^+_2={\lambda+\sqrt{\lambda^2+4\lambda}\over
2}.\eqno(3.10)
$$
Using this equality from first equation of the system (3.8) we
have for $k=2$
$$
(z^-_1)^{(1)}=\bigg({1+\sqrt{1-4a^2}\over 2a}\bigg)^2, \ \
(z^-_1)^{(2)}=\bigg({1-\sqrt{1-4a^2}\over 2a}\bigg)^2, \eqno(3.11)
$$ if $\l>{9\over 4}$ where $a=2(\sqrt{\lambda}+\sqrt{\lambda+4})^{-1}.$
Using $(z^-_1)^{(i)}, i=1,2$ and (3.10) we get
$(z^+_2)^{(1)}=\bigg({1-\sqrt{1-4a^2}\over 2a}\bigg)^2$,
$(z^+_2)^{(2)}=\bigg({1+\sqrt{1-4a^2}\over 2a}\bigg)^2.$ Similarly
from the second and third equality of (3.7) we get
$$z^+_1\in M=\left\{\bigg({1+\sqrt{1-4a^2}\over 2a}\bigg)^2,
\bigg({1-\sqrt{1-4a^2}\over 2a}\bigg)^2\right\}, \ \ z^-_2\in M.$$
Note that $(z^{\pm}_i)^{(1)}={1\over (z^{\mp}_i)^{(2)}}, \ \ i=1,2.$

Thus we proved \vskip 0,5 truecm

{\bf Proposition 3.} {\it If $k=2$ then for the case hinge}

1) {\it for $\lambda\leq {9\over 4}$ system (3.7) has unique
solution $z^*$;}

2) {\it for $\lambda> {9\over 4}$ system (3.7) has three solutions
$z^*_1=(z^-, {1\over z^-}, z^-, {1\over z^-})$, $z^*_2=({1\over
z^-}, {1\over z^-}, z^-, z^-)$, $z^*_3=(z^-, z^-, {1\over z^-},
{1\over z^-})$ where $z^-=\bigg({1-\sqrt{1-4a^2}\over 2a}\bigg)^2.$}
\vskip 0,5 truecm

Note that for $\l>{9\over 4}$ we have $0<a<{1\over 2}$ and $z^-<1.$

{\bf Corollary 2.} {\it If $k=2, \l>{9\over 4}$ then for any
solution of (3.1) (case hinge) we have  $z^-\leq z_{i,x}\leq {1\over
z^-},$ $i=1,2.$}\vskip 0,3 truecm

{\bf Remark 4.} To get exact solutions of (3.7) for $k=2$ we used
independence of first and last equations of (3.7) from the second
and third ones. But there is not such an independence for the pipe
and wrench case. Thus an analogue of Corollary 2 is not clear for
these cases.

\subsection{Case wand.} In this case from (3.1) for $z_x=z$ we have
$$ \left\{\begin{array}{ll}
z_1=\lambda\bigg({1+ z_1\over z_1+z_2}\bigg)^k,\\
z_2=\lambda\bigg({1+z_2\over z_1+z_2}\bigg)^k.
\end{array}\right.\eqno(3.12)$$

This case is very similar to the case hinge and one can prove that
if $k=2, \ \ \l>1$ then the system (3.12) has three solutions given
by similar formulas of case hinge just replacing $a$ with
$a=2(\sqrt{\l}+\sqrt{\l+8})^{-1}.$

Thus one can formulate an analogue of Theorem 2 with $\l_{cr}=1.$
But we have not analogues of Propositions 1-3 for the case wand.

\subsection{Case pipe}
In this case from (3.1) for $z_x=z$ we have
$$ \left\{\begin{array}{ll}
z_1=\lambda\bigg({1+ z_2\over 1+z_1}\bigg)^k,\\
z_2=\lambda\bigg({z_1\over 1+z_1}\bigg)^k.
\end{array}\right.\eqno(3.13)$$
From this we get $(x=z_2)$
$$\l^{-1}x=f(x)=\bigg({\sqrt[k+1]{x(1+x)^k}\over
1+\sqrt[k+1]{x(1+x)^k}}\bigg)^k. \eqno(3.14)$$ We have
$$f'(x)={k\over k+1}\cdot {(k+1)x+1\over x(x+1)}\cdot {(\sqrt[k+1]{x(1+x)^k})^k\over
(1+\sqrt[k+1]{x(1+x)^k})^{k+1}}>0. $$

Note that the equation (3.14) has at least one positive solution,
since $f$ is increasing and $f(0)=0, f(+\infty)=1.$ It is easy to
see that equation (3.14) has more than one positive solution if and
only if there is more than positive solution to $xf'(x)=f(x),$ which
is the same as
$$(k^2-1)x=\varphi(x)=(k+1)(1+x)\sqrt[k+1]{x(1+x)^k}+1.
\eqno(3.15)$$

Repeating this argument one can see that (3.15) has more than one
solution if and only if such is $x\varphi'(x)=\varphi(x)$. This
equation has the form
$$(k+1)x=\psi(x)={1\over \sqrt[k+1]{x(1+x)^k}}+k. \eqno(3.16)$$
Since the function $\psi(x)$ is decreasing the equation (3.16) has
unique solution. Consequently the system (3.13) has unique solution.

Thus we have proved

{\bf Theorem 3.} {\it For the case pipe  $\forall \l>0,\ \ \forall
k\geq 1,$ the translation-invariant pipe-hard core Gibbs measure is
unique.}\\[2mm]

{\bf Remark 5.} For $k=2$ this theorem was proved in [7].

For the case pipe one can prove the following propositions which are
analogues of Propositions 1 and 2.

{\bf Proposition 4.} {\it If $z_x=(z_{1,x}, z_{2,x})$ is a solution
of (3.1) in the case pipe then $z^-_i\leq z_{i,x} \leq z^+_i, \ \
\mbox{for any} \ \ i=1, 2, \ \ x\in V,$ where $(z^-_1, z^+_1, z^-_2,
z^+_2)$ is a solution of}
$$
\left\{\begin{array}{llll}
z^-_1=\lambda\bigg({1+z^-_2\over 1+z^+_1}\bigg)^k,\\
z^+_1=\lambda\bigg({1+z^+_2\over 1+z^-_1}\bigg)^k,\\
z^-_2=\lambda\bigg({z^-_1\over 1+z^-_1}\bigg)^k,\\
z^+_2=\lambda\bigg({z^+_1\over 1+z^+_1}\bigg)^k.
\end{array}\right.\eqno(3.17)
$$

\vskip 0,3 truecm

{\bf Proposition 5.} {\it If $z=(z_1^-, z^+_1, z^-_2, z^+_2)$ a
solution of (3.17) then $z^-_1=z^+_1$ iff $z^-_2=z^+_2$.} \vskip 0,3
truecm

{\bf Remark 6.} 1) For the case pipe we have not an analogue of
Proposition 3 and Corollary 2 since in this case there is no an
independence (mentioned in Remark 4) between equations of (3.17).

2) Next two sections are devoted to description of periodic and some
non-periodic Gibbs measures for the case hinge. Results of these
sections can be similarly proved for case pipe. But for the case
wand one needs to prove an analogue of Proposition 2.

\section{ Description of periodic Gibbs measures: case hinge}

\qquad For the case hinge  we write (3.1) in the following form
$$
h_{1,x}=\ln\lambda +\sum_{y\in S(x)}\ln{1+ \exp(h_{1,y})\over
1+\exp(h_{1,y})+\exp(h_{2,y})} ,
$$
$$
h_{2,x}=\ln\lambda+ \sum_{y\in S(x)}\ln{1+\exp(h_{2,y})\over
1+\exp(h_{1,y})+\exp(h_{2,y})}, \eqno(4.1)
$$
where $h_{i,x}=\ln z_{i,x}, \ \ i=1,2$. In this section we study
periodic  solutions of system (4.1).

 Note that (see [4]) there exists a one-to-one
correspondence between the set  $V$ of vertexes of the Cayley tree
of order $k\geq 1$ and the group $G_{k}$ of the free products of
$k+1$ cyclic  groups  of the second order with generators
$a_1,a_2,...,a_{k+1}$.

{\bf Definition 3.} Let $H_0$ be a subgroup of $G_k$. We say that
a collection $h=\{h_x=(h_{1,x}, h_{2,x}) : x\in G_k\}$ is {\it
$H_0$-periodic} if $h_{i,yx}=h_{i,x}$ for all $i=1,2$, $x\in G_k$
and $y\in H_0$.

{\bf Definition 4.} A Gibbs measure is called {\it $H_0$-periodic}
if it corresponds to an $H_0$-periodic collection $h$.

Observe that a translation-invariant Gibbs measure is
$G_k$-periodic.

Define function $h=(h_1,h_2)\mapsto F(h)=(F_1(h), F_2(h))$ where
$$
 F_1(h)=\ln{1+\exp(h_1) \over
1+\exp(h_1)+\exp(h_2)},\ \ F_2(h)=\ln{1+\exp(h_2) \over
1+\exp(h_1)+\exp(h_2)}. \eqno(4.2)
$$\vskip 0,3 truecm

{\bf Proposition 6.} {\it  $F(h)=F(l)$ if and only if $h=l$.} \vskip
0,3 truecm

{\bf Proof.} {\sl Necessity}. Let $F(h)=F(l)$ then $F_1(h)=F_1(l)$,
$F_2(h)=F_2(l)$, where $h=(h_1,h_2)$, $l=(l_1,l_2)$. From this
equalities we obtain
$$
\left\{\begin{array}{ll}
-t_2(z_1-t_1)+(1+t_1)(z_2-t_2)=0,\\
(1+t_2)(z_1-t_1)-t_1(z_2-t_2)=0,\\
\end{array}\right.\eqno(4.3)
$$
where $z_i=\exp(h_i),\ \  t_i=\exp(l_i), \ \ i=1,2.$ Note that
determinant of system (4.3) is negative, i.e.
$\Delta=-(1+t_1+t_2)<0.$ Therefore, (4.3) has unique solution
$z_i=t_i,\ \ i=1,2.$

{\sl Sufficiency}. Straightforward.

Let $G^2_k$ be the subgroup in $G_k$ consisting of all words of
even length. Clearly, $G^2_k$ is a subgroup of index 2. $H_0$ be a
normal subgroup of finite index in $G_k.$ We put $I(H_0)=H_0\cap
\{a_1,...,a_{k+1}\},$ where $a_i, \ \ i=1,...,k+1$ are generators
of $G_k$.

Using a similar argument of [7] one can prove following theorems.
\vskip 0,5 truecm

{\bf Theorem 4.} {\it For any normal subgroup of finite index each
$H_0-$ periodic Gibbs measure of hinge-hard core model is either
translation-invariant or $G^2_k-$periodic.} \vskip 0,5 truecm

{\bf Theorem 5.} {\it If $I(H_0)\ne \emptyset$ then each $H_0-$
periodic Gibbs measure   is translation-invariant.} \vskip 0,5
truecm

Theorems 4 and 5  reduce the problem of describing $H_0-$ periodic
Gibbs measure  with $I(H_0)\ne \emptyset$ to describing the fixed
points of the map $h=(h_1,h_2)\to (\ln\lambda,\ln\lambda)+kF(h),$
which describes translation-invariant Gibbs measures. If
$I(H_0)=\emptyset$, this problem is reduced to describing the
solutions of the system:
$$
\left\{\begin{array}{ll}
h=(\ln\lambda,\ln\lambda)+kF(l),\\
l=(\ln\lambda,\ln\lambda)+kF(h).\\
\end{array}\right.\eqno(4.4)
$$

Note that system (4.4) describes of periodic measures with period
two, precisely, $G^2_k-$ periodic measures.

Recall $z_i=\exp(h_i),\ \  t_i=\exp(l_i), \ \ i=1,2.$ Then  from
(4.4) we get
$$
\left\{\begin{array}{llll}
z_1=\lambda\bigg({1+t_1\over 1+t_1+t_2}\bigg)^k,\\
z_2=\lambda\bigg({1+t_2\over 1+t_1+t_2}\bigg)^k,\\
t_1=\lambda\bigg({1+z_1 \over 1+z_1+z_2}\bigg)^k,\\
t_2=\lambda\bigg({1+z_2\over 1+z_1+z_2}\bigg)^k.\\
\end{array}\right.\eqno(4.5)
$$

The analysis of solutions to  system  (4.5) is rather tricky. In a
particular case we shall reduce the  system  (4.5) to equation
$\gamma(\gamma(x))=x$ for some function $\gamma$ and will apply the
following lemma.

{\bf Lemma 1.} (See [6]) {\it Let $f:[0,1]\to [0,1]$ be a continuous
function with a fixed point $\xi\in (0,1).$ Assume that $f$ is
differentiable at $\xi$ and that $f'(\xi)<-1.$ Then there exist
$x_0,$ $x_1,$ $0 \leq x_0 < \xi < x_1 \leq 1,$ such that
$f(x_0)=x_1$ and $f(x_1)=x_0.$}\\[2mm]

If $z_1=z_2=z$, $t_1=t_2=t$ then (4.5) reduces to following system
$$
\left\{\begin{array}{ll}
z=\lambda\bigg({1+t\over 1+2t}\bigg)^k,\\
t=\lambda\bigg({1+z\over 1+2z}\bigg)^k.
\end{array}\right.\eqno(4.6)
$$
Denote $\gamma(x)=\lambda\bigg({1+x\over 1+2x}\bigg)^k.$ Then from
(4.6) we have $$z=\gamma(t),\ \ t=\gamma(z).\eqno(4.7)$$ \vskip 0,3
truecm

Note that the equation $x=\gamma(x)$ has unique solution
$x^*=x^*(k,\l),$ for any $k\geq 1$ and $\l>0.$

{\bf Theorem 6.} {\it For $k\geq 6$ and
$$
\lambda\in \left\{\lambda: {{k-3-\sqrt{(k-3)^2-8}}\over
4}<x^*<{{k-3+\sqrt{(k-3)^2-8}}\over4}\right\} \eqno(4.8)
$$
there are three $G^2_k$-periodic measures $\mu_0,$ $\mu_*,$ $
\mu_1,$. Which corresponds to three solutions $(x_0,x_1),$ $
(x_*,x_*),$ $(x_1,x_0)$ of (4.7).}\vskip 0,5 truecm

{\bf Proof.}  Note that function $\gamma(x)$ is decreasing for any
$x>0$. By Lemma 1, if $x^*$ satisfies
$$
\left\{\begin{array}{ll}
\gamma(x^*)=x^*,\\
\gamma'(x^*)<-1,\\
\end{array}\right.\eqno(4.9)
$$
then (4.6) has two solutions. From (4.9) it follows that
$$2(x^*)^2+(3-k)x^*+1<0. \eqno(4.10)$$
Solving this inequality we get $k\geq 6$ and (4.8). $\blacksquare$

{\bf Remark 7.} Theorem 6 gives more applicable conditions (for the
case hinge) than Theorem 4 of [7] (for case wrench).

\section{ Non-Periodic Gibbs measures: case hinge}

\qquad As follows from general results of [12,13], if a periodic
Gibbs measure is non-unique then there exist at least countable many
non-periodic Gibbs measures. This made more precise in theorem 8
below (for case hinge).

 We will show that system of equations (4.1) admit uncountably many
non-translational-invariant solutions. Take an arbitrary infinite
path $\pi=\{x^0=x_0<x_1<x_2<...\}$ on the Cayley tree starting at
the origin $x_0=x^0$. Establish a 1-1 correspondence between such
paths and real numbers $t\in [0,1]$ ([8], [9]). Write $\pi=\pi(t)$
when it is desirable to stress the dependence upon $t$. Map path
$\pi$ to a function $h^{\pi}:\;x\in V\mapsto h_x^{\pi}$ satisfying
(4.1). Note that $\pi$ splits Cayley tree $T^k$ into two subgraphs
$T_1^k$ and $T_2^k$.

For $k=2, \l>{9\over 4}$ the function $h^{\pi}$ is  defined by
$$h_{x}^{\pi}=\left\{\begin{array}{ll}
\ln z^-, \ \ \mbox{if} \ \ x\in T^k_1,\\
-\ln z^-, \ \ \mbox{if}\ \  x\in T^k_2,\\
\end{array}\right.\eqno (5.1)$$
where $z^-=\bigg({1-\sqrt{1-4a^2}\over 2a}\bigg)^2$, (see
Proposition 3).

Let $h\mapsto F(h)$ be defined by (4.2).\vskip 0,5 truecm

{\bf Proposition 6.} {\it For $k=2,$ $\l>{9\over 4}$ and  any
$h=(h_1,h_2)\in [\ln z^-; -\ln z^-]^2$ (recall $z^-<1)$ the
following inequalities hold:}

a) $$ \left|{\partial F_1\over \partial h_1}\right| \leq {1\over
(\sqrt{z^-+1}+\sqrt{z^-})^2}; \ \ \left|{\partial F_2\over \partial
h_2}\right| \leq {1\over (\sqrt{z^-+1}+\sqrt{z^-})^2};
$$
$$
\left|{\partial F_1\over \partial h_2}\right| \leq {1\over
1+z^-+(z^-)^2}; \ \ \left|{\partial F_2\over \partial h_1}\right|
\leq {1\over 1+z^-+(z^-)^2}; $$

b) $$ \|F(h)-F(l)\|\leq {2\over 1+z^-+(z^-)^2} \|h-l\|.$$ \vskip 0,5
truecm

{\bf Proof.} a) Using lemma 9 of [11] and $\ln z^-\leq h_i\leq -\ln
z^-, \ \ i=1,2$  we have
$$\left|{\partial F_1\over \partial h_1}\right| \leq {\exp(h_2)\over
(\sqrt{\exp(h_2)+1}+1)^2}=\psi(h_2).$$ The function $\psi(x)$ is
increasing, therefore
$$\left|{\partial F_1\over \partial h_1}\right| \leq
\psi(-\ln{z^-})={1\over (\sqrt{z^-+1}+\sqrt{z^-})^2}.$$ The proof
for $\left|{\partial F_2\over
\partial h_2}\right| $ is similar.

Now consider $$\left|{\partial F_1\over \partial h_2}\right|=
{\exp(h_2)\over 1+\exp(h_1)+\exp(h_2)}=\varphi(h_1, h_2).$$ Since
$\varphi^{'}_{h_{2}}>0$ and $\varphi^{'}_{h_{1}}<0$ we have
$$\left|{\partial F_1\over \partial h_2}\right| \leq
\max\varphi(h_1,h_2)=\varphi(\ln{z^-}, -\ln{z^-})={1\over
{1+z^-+(z^-)^2}}$$ Similarly one can show that
$$\left|{\partial F_2\over
\partial h_1}\right| \leq {1\over
1+z^-+(z^-)^2}.$$

b) For $z^-<1$ it is easy to see that  ${1\over
(\sqrt{z^-+1}+\sqrt{z^-})^2}<{1\over 1+z^-+(z^-)^2}.$ Using this
inequality we obtain

$$
\|F(h)-F(l)\|=\max_{i=1,2}\{|F_i(h)-F_i(l)|\}\leq$$
$$\max_{i=1,2}\{|(F_i)'_{h_1}||h_1-l_1|+|(F_i)'_{h_2}||h_2-l_2|\}\leq
 {2 \over 1+z^-+(z^-)^2}\|h-l\|.$$ This completes the proof.

If ${2\over 1+z^-+(z^-)^2}<1$ i.e. $z^->{\sqrt{5}-1\over 2}$ then
with the help of Proposition 6 it is easy to prove the following
Theorem 7, similar to Theorem 3 of [9]: \vskip 0,5 truecm

{\bf Theorem 7.} {\it If $k=2$, and $\l$ is such that
${\sqrt{5}-1\over 2}<z^-<1$ then for any infinite path $\pi$ there
exists a unique function $h^{\pi}$ satisfying} (4.1) {\it  and}
(5.1). \vskip 0,3 truecm

In the standard way (see [4], [9], [12]) one can prove that
functions $h^{\pi(t)}$ are different for different  $t \in [0;1].$
Now let $\mu(t)$ denote the Gibbs measure corresponding to function
$h^{\pi(t)}$, $t\in [0;1].$ Similarly to theorem 3.2 from [1], we
can prove the following: \vskip 0,3 truecm

{\bf Theorem 8.} {\it If conditions of Theorem 7 are satisfied then
for any  $t\in [0;1]$, there exists a unique hinge-hard core Gibbs
measure $\mu(t)$. Moreover, the Gibbs measures $\mu_1$, $\mu_2$ (see
Theorem 2) are specified as} $\mu(0)=\mu_1$ and
$\mu(1)=\mu_2.$\vskip 0,2 truecm

Because measures $\mu(t)$ are different for different $t\in [0,1]$
we obtain a continuum of distinct Gibbs measures which are
non-periodic.

 \vskip 0.2 truecm

 {\bf Acknowledgments.}  A part of this work was done
within the scheme of Junior Associate at the ICTP, Trieste, Italy
and the first author (UAR) thanks ICTP for providing financial
support and all facilities (in May - August 2006).  He also thanks
the IHES, Bures-sur-Yvette, France for support and kind hospitality
(in October - December 2006) and the SMS, Lahore, Pakistan where
final part of this work was done.

\vskip 0.2 truecm

\textbf{References}

1. Bleher P.M. and  Ganikhodjaev N.N., On pure phases of the Ising
model on the Bethe lattice. {\it Theor. Probab. Appl.} {\bf 35}
(1990), 216-227.

2. Brightwell G. and  Winkler P., Graph homomorphisms and phase
transitions. {\it J. Combin. Theor.} Series B, {\bf 77}, (1999),
221-262.

3. Brightwell G., H\"aggstr\"om O. and  Winkler P., Non monotonic
behavior in hard-core and Widom-Rowlinson models. {\it J. Statist.
Phys.}, {\bf 94}, (1999), 415-435.

4. Ganikhodjaev N. N. and Rozikov U. A., Description of periodic
extreme Gibbs measures of some lattice models on the Cayley tree.
{\it Theor. Math. Phys}.{\bf 111} (1997), 480-486.

5. Georgii H.-O., Gibbs measures and phase transitions. \textit{De
Gruyter studies in math.} {\bf 9}, Berlin, New-York, 1988.

6. Kesten H., Quadratic transformations: A model for population
growth. I. {\it Adv. Appl. Prob.} {\bf 2}, (1970), 1-82.

7. Martin J., Rozikov U.A., Suhov Yu.M., A Three state Hard-Core
model on a Cayley tree, {\it J. Nonlinear Math. Phys.}, {\bf 12:3},
(2005), 432-448.

8. Rozikov U. A., Description of limiting Gibbs measures for
$\lambda-$ models on the Bethe lattice, {\it Siberian Math. Journ.}
{\bf 39} (1998), 427-435.

9. Rozikov U. A., Description uncountable number of Gibbs measures
for inhomogeneous Ising model. {\it Theor. Math. Phys.}{\bf 118}
(1999), 95-104.

10. Rozikov U.A., Suhov Yu.M., A hard-core model on a Cayley tree:
an example of a loss network, {\it Queueing Syst.}, {\bf 46},
(2004), 197-212.

11. Rozikov U. A., Shoyusupov Sh. A., Gibbs measures for the SOS
model with four states on a Cayley tree. {\it Theor. Math.
Phys.}{\bf 149:1} (2006), 1312-1323.

12. Zachary S., Countable state space Markov random fields and
Markov chains on trees. {\it Ann. Prob.} {\bf 11} (1983), 894--903.

13. Zachary S., Bounded, attractive and repulsive Markov
specifications on trees and on the one-dimensional lattice. {\it
Stochastic Process. Appl.} {\bf 20} (1985), 247--256.

\end{document}